# Experimental and theoretical investigations on magnetic and related properties of ErRuSi

Sachin Gupta,[1] A. Das,[2] K. G. Suresh,[1,*] A. Hoser,[3] Yu.V. Knyazev,[4] Yu. I. Kuz'min,[4] and A. V. Lukoyanov[4,5]

[1]Department of Physics, Indian Institute of Technology Bombay, Mumbai-400076, India

[2]Solid State Physics Division, Bhabha Atomic Research Centre, Mumbai-400085, India

[3]Helmholtz-Zentrum Berlin, Hahn-Meitner Platz 1, 14-109 Berlin, Germany

[4]Institute of Metal Physics, Russian Academy of Sciences, Ural Branch, Yekaterinburg - 620990, Russia

[5]Ural Federal University, Yekaterinburg - 620002, Russia

## Abstract

We report experimental and theoretical studies of magnetic and related properties of ErRuSi compound. Various experimental techniques such as neutron diffraction, magnetization, magneto-thermal, magneto-transport, optical have been used to study the compound. Neutron diffraction shows ferromagnetic ordering at low temperatures with moments aligned in *ab* plane. Neutron diffraction and magnetization data show reduction in magnetic moment, which may be due to crystalline electric field effects at low temperatures. The compound shows good magnetocaloric properties with a low field adiabatic temperature change of 4.7 K, which is larger than that of many proposed materials for magnetic refrigeration at low temperatures. Magnetoresistance shows large negative value at 8 K, which changes its sign and increases in magnitude, with decrease in temperature and/or increase in field. The positive MR at low temperatures attributed to the Lorentz force effect. The electronic structure calculations accounting for electronic correlations of the 4f electrons of Er reproduces the ferromagnetic ordering and effective magnetic moment. Interband transitions between the Ru and Er d states and Er f states in one spin projection are found to form the main features of the measured optical conductivity in this compound.

# 1. Introduction

In rare earth intermetallic compounds, RTX (R= rare earth, T=transition metal and X= p-block elements) series is an extensively studied series of compounds owing to its fascinating physical properties [1]. These compounds also show good hydrogen storage capacity, large magnetocaloric effect (MCE) and large magnetoresistance (MR), which make these compounds suitable for potential applications. It has been observed that except Mn, no other transition element in RTX contributes significant moment. Therefore, in such compounds, only rare earth ion contributes to the net moment. Many compounds of this series are found to show interesting magnetic and electrical properties [1]. One of the interesting series in the RTX family is RRuSi, where Ru is the 4d element. It has been found that magnetic properties of RRuSi series vary considerably as R is varied. Welter et al. [2] performed neutron diffraction and magnetic measurements on some RRuSi (R=La-Nd, Sm, Gd) compounds. The authors observed that all these compounds crystallize in CeFeSi type tetragonal structure (space group: *P4nmm*). The report [2] shows that LaRuSi is Pauli paramagnet, CeRuSi is Curie-Weiss paramagnetic down to 4.2 K, PrRuSi and NdRuSi are antiferromagnets, while SmRuSi and GdRuSi are ferromagnets. Focusing on some of the compounds of this series, recently some of us have studied magnetic and magnetocaloric properties of ErRuSi compound [3]. It has been observed that this compound crystallizes in the orthorhombic crystal structure with space group *Pnma*. Magnetic properties show that it orders ferromagnetically below 8 K [3]. The observed effective magnetic moment ($\mu_{eff.}$=9.48 $\mu_B$/Er$^{3+}$) and the moment at the lowest measured temperature for the highest measured magnetic field ($M_{sat.}$=6.94 $\mu_B$/f.u.) estimated from the magnetization data ($\mu_{th.}$=9.59 $\mu_B$/Er$^{3+}$, gJ= 9 $\mu_B$/Er$^{3+}$) are found to be smaller than the theoretical values (the case is similar to NdRuSi). This was earlier attributed to a small Ru moment that may couple antiferrromgnetically with Er moment [3]. To confirm these predictions, neutron diffraction measurements were needed. The compound was found to show giant magnetocaloric effect (GMCE) along with negligible thermal and field hysteresis [3]. Generally it has been observed that the compound with large MCE shows large MR as well [4]. It is attributed to the change in magnetic structure on the application of field, which affects both magnetocaloric and magneto-transport properties.

In this paper we have studied ErRuSi compound in detail by means of various experimental probes such as neutron diffraction, magnetization, optical, magneto-thermal and magneto-transport measurements. The electronic band structure calculations were also performed to support the experimental results.

## 2. Experimental and computational details

The polycrystalline sample of ErRuSi was synthesized by the arc melting method taking stoichiometric amount of its constituent elements. The melted sample was sealed in evacuated quartz tube and annealed for 7 days at 800 ºC followed by furnace cooling. The magnetization, M (T, H), and the heat capacity C (T, H) were carried out on Physical Property Measurement System (Quantum Design). The resistivity/magnetoresistance measurements were performed in homemade setup along with 8-Tesla Oxford superconducting magnet system in longitudinal geometry. For neutron diffraction measurements 6 g of sample was prepared and crushed to make fine powder. The neutron diffraction measurements at different temperatures and zero field have been carried out on E6 diffractometer at Helmholtz Zentrum Berlin, Germany.

The electronic structure of ErRuSi was obtained within the LSDA+U method [5] that combines the local spin density approximation (LSDA) and Hubbard U correction of electronic correlations in the 4$f$ shell of erbium. The TB-LMTO-ASA package [6] was used; it is based on the linear muffin-tin orbitals with atomic sphere and tight binding approximations. The following muffin-tin orbitals were included into the orbital basis set: (6$s$, 6$p$, 5$d$, 4$f$) states of Ho, (5$s$, 5$p$, 4$d$, 4$f$) states of Rh, and (3$s$, 3$p$, 3$d$) states of Si. Radii of the muffin-tin orbitals were R(Er) = 3.63 a.u., R(Ru) = 2.69 a.u. and R(Si) = 2.55 a.u. The values of the Coulomb U = 8 eV and exchange Hund $J_H$ = 0.7 eV parameters for the 4$f$ shell of Er are close to the ones used in previous calculations [7,8]. These values were used to account for strong correlations of the 4f electrons of erbium in the LSDA+U calculations. The ferromagnetic ordering of the Er magnetic moments was modeled.

## 3. Results and discussion
### 3.1 Neutron diffraction

The neutron diffraction pattern was recorded at several temperatures between 1.6 and 25 K for ErRuSi compound. Fig. 1 shows the Rietveld refinement of neutron diffraction pattern recorded at 9 and 1.6 K. Analysis of neutron diffraction data at 9 K shows paramagnetic behavior of ErRuSi. The Rietveld analysis in the paramagnetic regime confirms orthorhombic crystal structure (space group *Pnma*). In this structure, the Er, Ru and Si ions occupy the 4c (x 1/4 z), 4c (x 1/4 z) and 8d (x y z) positions, respectively. On lowering the temperature below 9K, strong enhancement in the intensity of the fundamental reflections (101), (002), (200) is found. No superlattice reflections was observed down to the lowest temperature. The absence of magnetic ordering in isostructural LaRuSi [2] indicates that the Ru ion does not carry a moment in this compound. Hence it is clear that similar to other compounds in the *RTX* series [1], i.e., in ErRuSi, only Er ion contributes to the magnetic moment. Therefore, the magnetic structure was modeled with moments only on the Er ion. The ferromagnetic structure obtained for ErRuSi is shown in Fig. 2. It can be noted from Fig. 2 that all the moments are parallel and aligned in the a-b plane, which confirms ferromagnetic ordering in this compound. A small canting of the moments away from the *b*-axis is found to fit the data better. The variation of Er moment with temperature is shown in Fig. 3(a). The Curie temperature estimated from the neutron diffraction data is 9 K, which is close to the value estimated from the magnetization data ($T_C$ =8 K). At 1.6 K, the magnitude of moment is 5.9 $\mu_B$ [see Fig. 3(a)], which is significantly lower than the theoretically estimated value.

Fig. 3(b) shows the temperature dependence of the unit cell volume, derived from the neutron data. It can be seen that the volume decreases with increase in temperature. The temperature dependence of magnetization for ErRuSi is shown in Fig. 4, while field dependence of magnetization is shown in the inset of Fig. 4. The $\mu_{eff}$ estimated from the Curie-Weiss fit to the magnetic susceptibility data in the paramagnetic regime is 9.48 $\mu_B/Er^{3+}$, which is slightly smaller than the theoretically expected value of free $Er^{3+}$ ($\mu_{th.}$= 9.59 $\mu_B/Er^{3+}$). The smaller value of $\mu_{eff}$ may be attributed to crystalline electric fields (CEF). The saturation moment estimated from the extrapolation of magnetization at 3 K is found to be 6.3 $\mu_B$, which is close to the value estimated from neutron diffraction at 1.6 K. The saturation moments observed from magnetization and neutron diffraction data are smaller than the theoretical saturation moment, (gJ=9 $\mu_B$). The assumption of small Ru moment made in earlier report is ruled out now in the light of the neutron diffraction data that shows no moment on Ru. Therefore, it is reasonable to assume that CEF plays a crucial role in determining the moment at low temperatures. Similar to this, some other compounds in *RTX* series also show smaller value of saturation moment [9, 10]. The magnetic parameters of the compounds of this series are given in Table I for comparison. It is clear from the table that light rare earth compounds of this series either are paramagnetic or show antiferromagnetic behavior while heavy rare earth compounds show ferromagnetic behavior with the exception of Sm compound, which also shows ferromagnetic behavior. The ordering temperature in ErRuSi is smaller than that of the other compounds of this series, which show antiferro or ferromagnetic ordering. It can also be noted from Table I that NdRuSi and GdRuSi show reduced moment. The authors reported that difference in observed and expected values of moment may arise due to CEF. However, the reduction in GdRuSi moment suggests that the reduction cannot be due to the first order contribution of CEF.

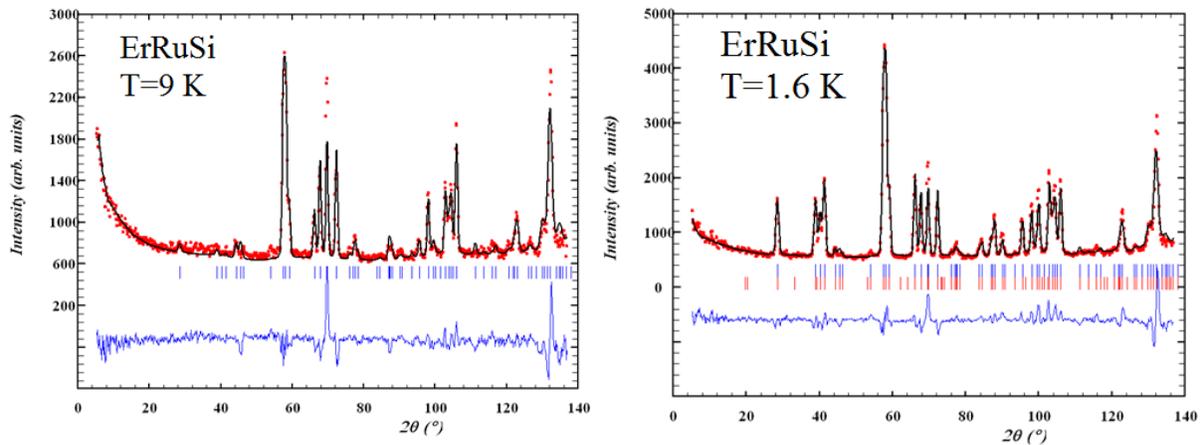

Fig. 1. Neutron diffraction patterns at 9 and 1.6 K for ErRuSi. The bottom lines show differences between observed and calculated intensities.

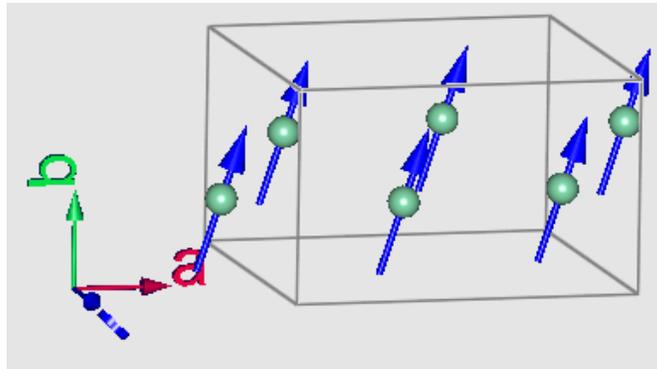

Fig. 2. Suggested magnetic structure for ErRuSi compound at 1.6 K.

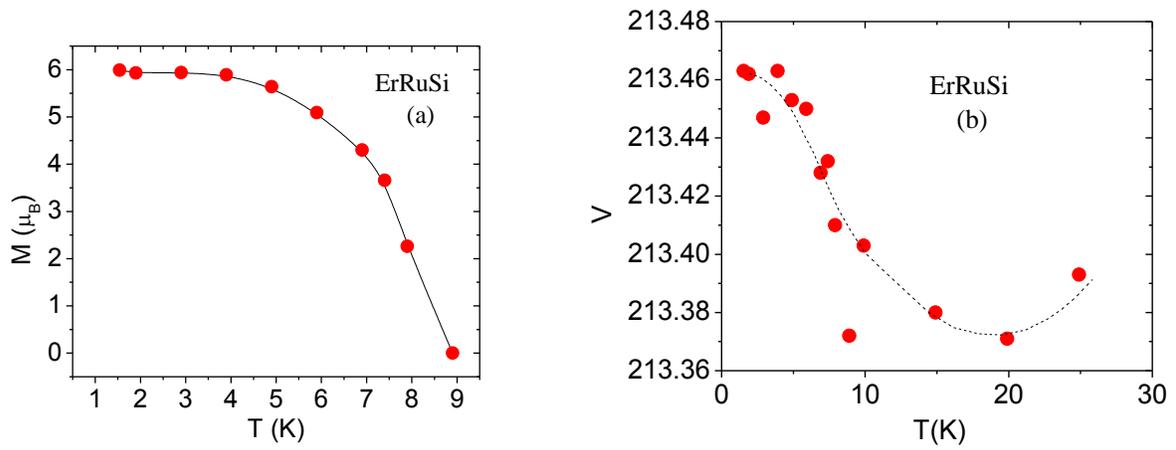

Fig. 3. The temperature evolution of (a) Er moment (b) unit cell volume for ErRuSi.

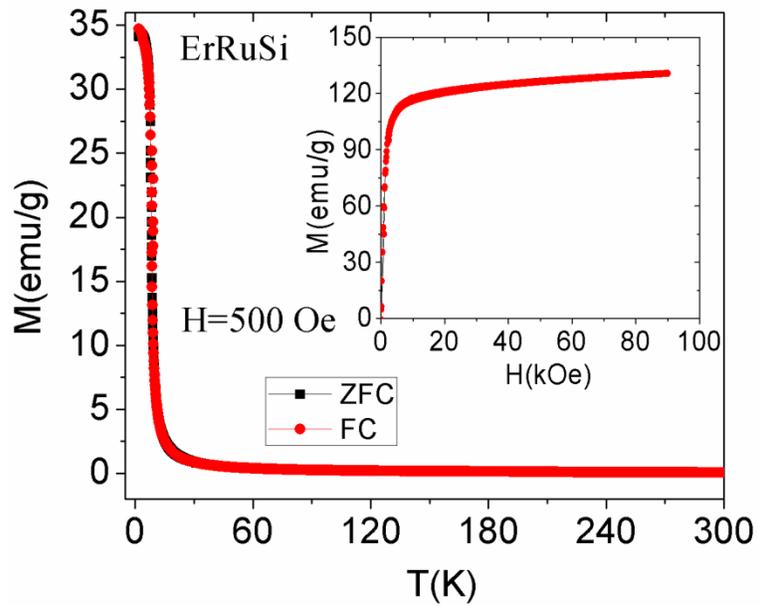

Fig. 4. The temperature dependence of magnetization for ErRuSi. The inset shows the field dependence of magnetization at 3 K.

**Table I: Magnetic parameters of members of RRuSi series.**

| Compound | Magnetic nature | $T_N$ or $T_C$ (K) | $\mu_{eff}$ ($\mu_B/R^{3+}$) | $g[J(J+1)]^{1/2}$ ($\mu_B$) | M ($\mu_B$) | gJ ($\mu_B$) | Ref. |
|---|---|---|---|---|---|---|---|
| LaRuSi | Pauli-paramagnetic | - | - | - | - | - | [2] |
| CeRuSi | Curie-Weiss paramagnetic | - | 2.56 | 2.54 | - | - | [2] |
| PrRuSi | antiferromagnetic | 73 | 3.52 | 3.58 | - | - | [2] |
| NdRuSi | antiferromagnetic | 74 | 3.46 | 3.62 | 2.8* | 3.2 | [2] |
| SmRuSi | ferromagnetic | 65 | NCW | NCW | 0.15 | 0.72 | [2] |
| GdRuSi | ferromagnetic | 85 | 8.58 | 7.94 | 6.4 | 7 | [2] |
| ErRuSi | ferromagnetic | 8 | 9.48 | 9.59 | 6.3, 9.59* | 9 | This study |

NCW= Non-Curie-Weiss, *= estimated from neutron diffraction

### 3.2 Heat capacity measurements

The heat capacity measurements in 0, 10, 20 kOe in the temperature range of 2-100 K is shown in Fig. 5. The zero field heat capacity shows λ-shaped peak around 8 K due to the onset of magnetic ordering. The peak gets suppressed and the peak position shifts to higher temperatures on the application of field, which confirms the ferromagnetic ordering in this compound.

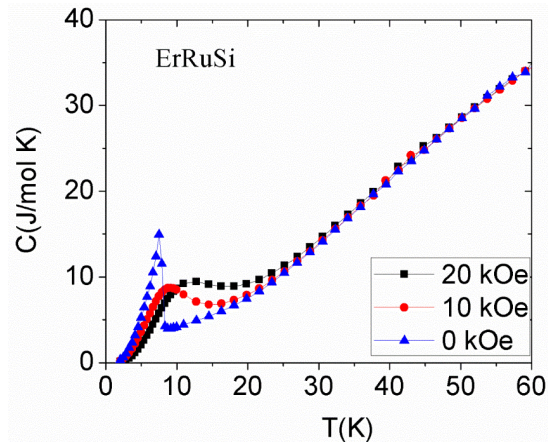

Fig. 5. The temperature dependence of the heat capacity of ErRuSi in 0, 10 and 20 kOe fields.

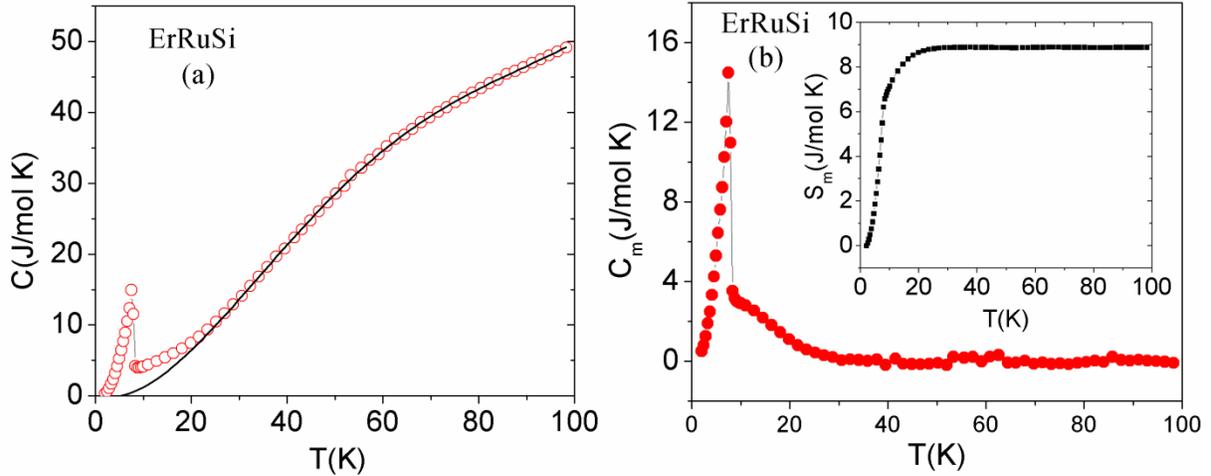

Fig. 6. (a) Non-magnetic fit to the total heat capacity data of ErRuSi (b) The temperature dependence of magnetic heat capacity. The inset shows the temperature dependence of total magnetic entropy.

To calculate the magnetic heat capacity, the non-magnetic part of the heat capacity (obtained by theoretical fitting and shown in Fig. 6 (a)) was subtracted from the total heat capacity. The total magnetic entropy ($S_m$) was calculated using the magnetic heat capacity (as shown in the main panel of Fig. 6(b)) employing the relation, $S_m = \int_0^T \frac{C_m}{T} dT$. The $S_m$ shows saturation at higher temperatures, as can be seen from the inset of Fig. 6 (b). At $T_C$, the value of $S_m$ is close to R ln2, which suggests that only one Kramer's doublet participates in the magnetic ordering. The saturation value of $S_m$ has been estimated to be 9 J/mol K. The expected value of $S_m$ is R ln (2J+1)=23 J/mol K. Therefore it is clear that approximately only 40% of moments takes part in magnetic ordering, which plausibly explains the low saturation moment in ErRuSi. It is reported earlier that the compound shows giant MCE around its ordering temperature. In view of this we calculated the change in adiabatic temperature ($\Delta T_{ad}$) from heat capacity data at 20 kOe field using the relation, $\Delta T_{ad}(T)_{\Delta H} \cong [T(S)_{H_f} - T(S)_{H_i}]_S$. The value of $\Delta T_{ad}$ at 20 kOe was found to be 4.7 K. The large, low field $\Delta T_{ad}$ along with large entropy change and the refrigerant capacity (RC, reported earlier) makes this compound promising for low temperature magnetic refrigeration.

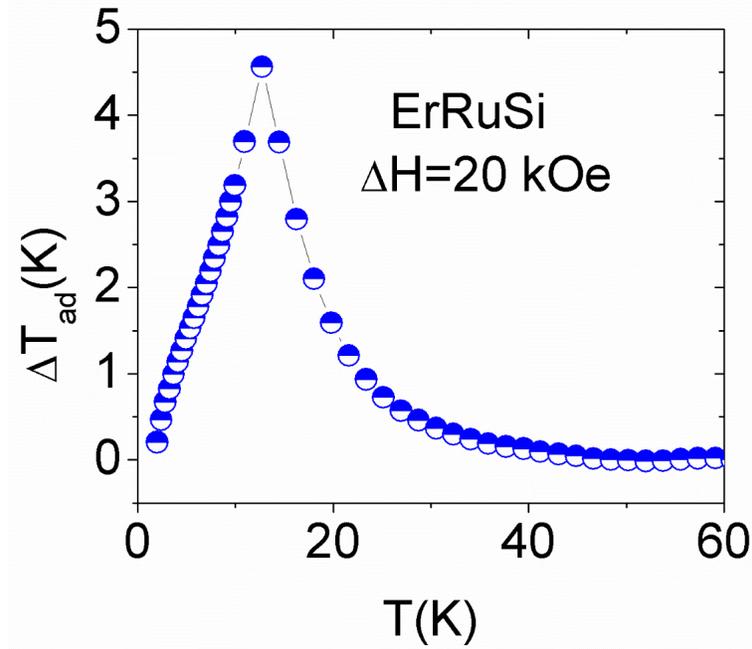

Fig. 7. The temperature dependence of adiabatic temperature change in ErRuSi for 20 kOe.

**3.3 Electrical resistivity and magneto-resistance (MR)**

The zero field electrical resistivity in the temperature range of 2-100 K is shown in Fig. 8(a). In paramagnetic regime the resistivity is almost linear showing an increase with temperature. At 8 K, the resistivity shows a slope change due to the ferromagnetic ordering in this compound. The $T_C$ estimated from the resistivity data is close to that found from the magnetization and the heat capacity plots. Fig. 8(b) shows the field dependence of MR at different temperatures. It can be noted from Fig. 8(b) that at 40 K, the MR is almost negligible, but negative. As the temperature decreases, the MR magnitude increases down to 8 K and field dependence gradually changes from nearly $H^2$ at 40 K to $\sqrt{H}$. Below 8 K, the MR is negligible at lower fields and becomes positive at higher fields. The magnitude of positive MR is about 4% at 1.5 K, for 50 kOe. In ferromagnetic materials, the carrier scattering due to magnons gets suppressed with increase in magnetic field, which results in negative magnetoresistance. The magnon contribution to resistivity decreases with lowering temperature below $T_C$. Therefore with lowering temperature the negative contribution to MR becomes less significant and in the present system it is almost negligible even at 4 K. The negligible magnon contribution below this temperature is consistent with the temperature variation of moment shown in figure 3(a), where

M is close to saturation magnetization for T≤4K. The positive MR at high fields, below $T_C$, arises due to the Lorentz force contribution, the magnitude of which increases with lowering temperature and/or increasing magnetic field.

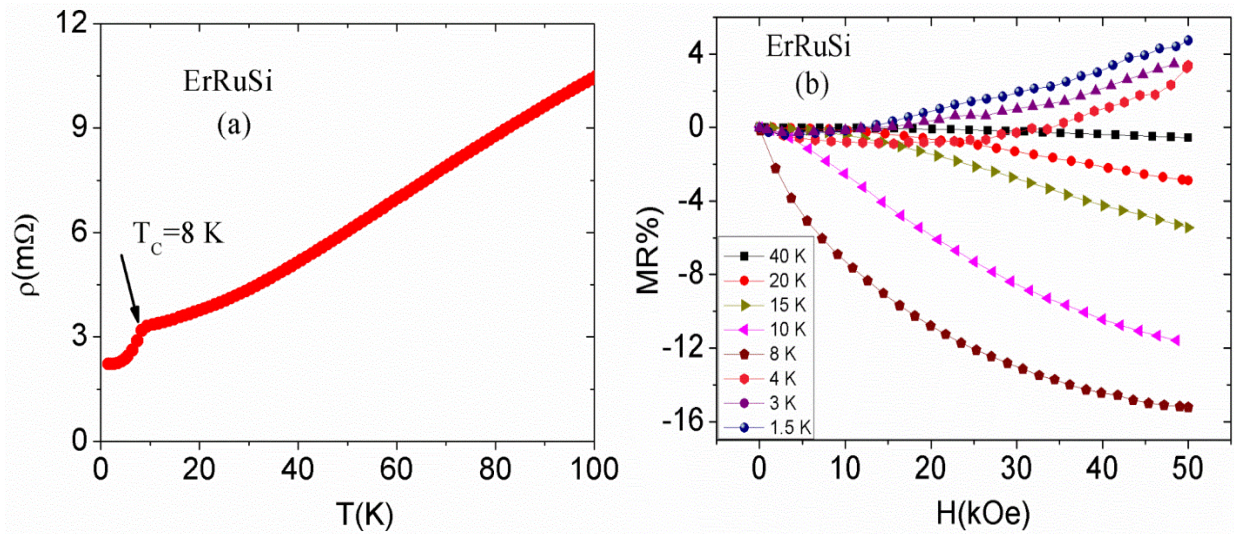

Fig. 8. (a) The temperature dependence of the zero field resistivity in ErRuSi. (b) The field dependence of MR at different temperatures.

### 3.4 Theoretical calculations and optical study

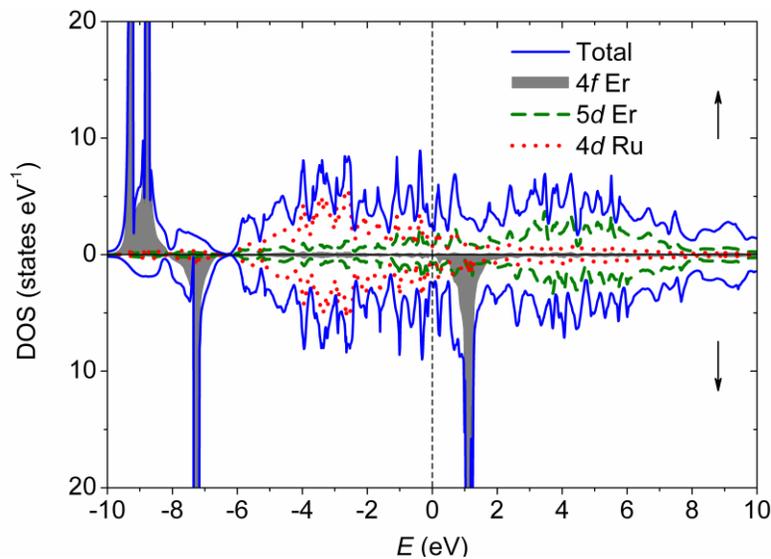

Fig. 9. The LSDA+U calculated total and partial densities of states of ErRuSi.

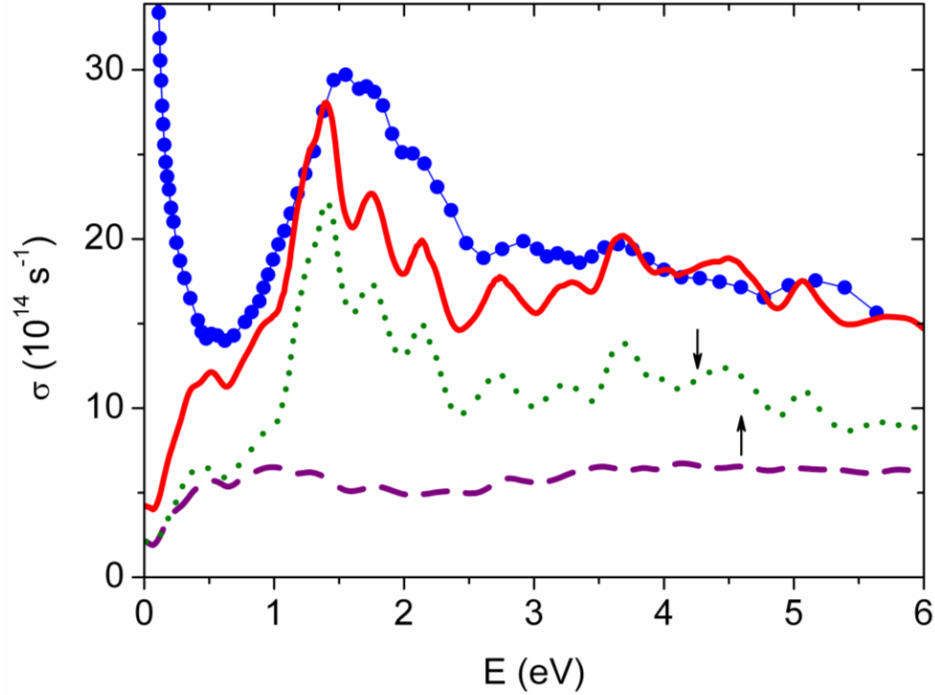

Fig. 10. The energy dependence of experimental (circles) and theoretical (solid line) optical conductivities in ErRuSi. Dashed and dotted lines show spin-up and spin-down contributions, correspondingly.

In the LSDA+U calculations for ErRuSi, a ferromagnetic ordering of magnetic moments of Er ions was obtained as the ground state. No magnetic moments were found on Rh and silicon ions in agreement with the neutron diffraction results. The Er moment was calculated as 3 $\mu_B$ per Er ion that includes spin moment but neglects orbital moment component. To account for the latter, one can employ the procedure adopted in ref. 11 and take into account L = 6 for the $Er^{3+}$ ion, then for the Lande factor g = 6/5, the effective magnetic moment of Er can be estimated as 9.5 $\mu_B$. This value is in good agreement with the experimental one, but gJ in this case is equal to 8.9 $\mu_B$ overestimating the saturation moment value that can be attributed to the neglect of the lattice changes at low temperatures or the simplified moment estimation procedure.

The total and partial densities of states obtained for ErRuSi in our LSDA+U calculation is shown in Fig. 9 relatively to the Fermi energy ($E_F$). The Er 4f states are centered around -9 eV for the filled bands in one spin projection, while for another one they are split into the filled part at -7.2 eV and the empty part at 1.5 eV above $E_F$. All other states are mostly without spin-polarization that results in magnetic moment of 0.05 $\mu_B$ for the Er 5d states and negligible moments of the others. Below the Fermi energy the extended occupied Ru 4d states contribute significantly from -6 eV, while the Er 5d are extended above $E_F$. The electronic states of silicon (not shown in Fig. 9) are hybridized with the other states and spread over the whole energy region.

Fig. 10 shows the optical conductivity $\sigma(E)$ for ErRuSi compound. This is the most sensitive spectral parameter that characterizes the energy dependence and intensity of optical response of medium. In the spectrum of $\sigma(E)$, two frequency ranges are well defined that correspond to two different types of electronic excitation by light: intra- and interband ones. This kind of behavior of the optical conductivity is typical for solids with metallic conductivity. In the low-energy infrared range (< 0.5 eV) a rapid increase in the optical conductivity is caused by the Drude mechanism of interaction of electromagnetic waves with electrons ($\sigma \sim \omega^{-2}$). In this spectral interval were estimated kinetic characteristics of conduction electrons, namely, plasma $\omega_p$ and relaxation $\gamma$ frequencies. Their values are frequency independent in wavelength range above 10 $\mu$m and equal to $\omega_p = 3.7 \cdot 10^{-15}$ s$^{-1}$ and $\gamma = 1.8 \cdot 10^{-14}$ s$^{-1}$. With the increase of light frequency (visible and ultraviolet intervals) quantum absorption starts to dominate. The spectrum of $\sigma(E)$ in this region reveals a broad intensive band with the peak at ~ 1.5 eV and some weak maxima above. These features are formed by interband transitions between the different states divided by the Fermi level and reflect the actual structure of electron spectrum of ErRuSi.

The electronic densities for the spin-up and spin-down states, see Fig. 9, were used to interpret the optical data. The interband optical conductivity was (shown in Fig. 10 in arbitrary units) calculated directly from the electronic structure through the convolution of the total DOS below and above $E_F$ in approximation that direct and indirect transitions are equally probable. The figure also demonstrates the partial contributions in conductivity from each of these spin subsystems. The calculated interband $\sigma(E)$ predict the strong absorption region at 0.7 – 2.4 eV that reproduces the experimental absorption band rather well. The maxima in this interval according to the calculations are mostly in the spin-down bands due to the electron transitions between the Ru 4d and Er 5d states below the $E_F$ and Er 4f states above $E_F$. For the higher photon energies structural nonmonotonies of $\sigma(E)$ curve, as it is seen from Fig. 10, are also formed by transitions in this spin subsystem. The contribution in conductivity from the electron transitions in opposite spin-up bands is rather small and nearly unchanged in the investigated range. On the whole experimental energy dependence of the optical conductivity for the ErRuSi in quantum absorption range is well described with the electronic structure obtained within the LSDA+U method.

## Conclusions

ErRuSi is ferromagnetic below 8 K, as confirmed by bulk magnetization and neutron diffraction data. Neutron data shows that Ru has no significant moment in this compound. The saturation moment is smaller than the theoretically expected value. The analysis of heat capacity data suggests that all the moments do not participate in magnetic ordering, which results in low value of saturation moment. Crystalline electric field effect may be the reason for smaller saturation moment. MR shows negative sign near ordering temperature and positive at low temperatures. The positive MR in this ferromagnetic material arises due to Lorentz force. The electronic structure calculations accounting for electronic correlations of the 4f electrons of Er reproduces the ferromagnetic ordering and effective magnetic moment. Using the calculated densities of states, interband transitions between the Ru and Er d states and Er f states in one spin projection are found to form the main features of the measured optical conductivity of ErRuSi.

## Acknowledgements

SG would like to thank CSIR, New Delhi and IIT Bombay for granting the fellowship. UGC-DAE Consortium for Scientific Research, Indore is acknowledged for providing Resistivity/MR facility. Dr. R Rawat and Sachin Kumar are acknowledged for RT/MR measurements. Theoretical calculations of the electronic structure were supported by the grant of the Russian Science Foundation (project no. 14-22-00004). Experimental optical measurements were supported by the Russian Foundation for Basic Research, project No. 13-02-00256a.